\documentclass[12pt]{article}
\usepackage{upgreek}
\usepackage{amsmath,amssymb,color}
\usepackage{amsfonts}
\usepackage[all]{xy}
\usepackage{graphics}

\newcommand{\ii}{{\rm i}}
\newcommand{\ep}{{\rm e}}
\newcommand{\be}{\begin{equation}}
\newcommand{\ee}{\end{equation}}
\newcommand{\bea}{\setlength\arraycolsep{2pt} \begin{eqnarray}}
\newcommand{\eea}{\end{eqnarray}}
\newcommand{\nn}{\nonumber}
\newcommand{\rd}{{\rm d}}

\renewcommand{\thefootnote}{\fnsymbol{footnote}}

\begin{document}
\begin{titlepage}
\vfill
\begin{flushright}

\end{flushright}
\vfill
\begin{center}
{\Large\bf Bulk Local Operators, Conformal Descendants \\ and Radial Quantization}

\vskip 1cm
Zhao-Long Wang$^{1,2}$\footnote{\tt zlwang4@gmail.com}, Yi Yan$^{1,2}$

\vskip 5mm
$^1${\it Institute of Modern Physics, Northwest University, XiAn 710069, China}
\\$^2${\it Shaanxi Key Laboratory for Theoretical Physics Frontiers, XiAn 710069, China}

\end{center}
\vfill

\begin{abstract}
\noindent
We establish a construction of the bulk local operators in AdS by considering CFT at finite energy scale.
Without assuming any prior knowledge about the bulk, the solution to the bulk free field equation automatically appears in the field theory arguments. In the radial quantization formalism, we find a properly regularized version of our initial construction.
Possible generalizations beyond pure AdS are also discussed.

\end{abstract}

\vfill
\end{titlepage}

\tableofcontents\newpage
\renewcommand{\thefootnote}{\#\arabic{footnote}}
\setcounter{footnote}{0}

\section{Introduction}
The AdS/CFT correspondence \cite{Maldacena:1997re,Gubser:1998bc,Witten:1998qj} implies a duality between the quantum gravity in $d+1$-dimensional anti-de Sitter space and the $d$-dimensional conformal field theory which is defined on the boundary of AdS$_{d+1}$. The AdS metric in the Poincar\'e patch is given by
\begin{eqnarray}\label{ads}
  {\rd s}^2 = \frac{1}{z^2}\left({{\rd z}^2}    + \eta_{\mu\nu}{\rd x}^{\mu}{\rd x}^{\nu}\right)\,,
\end{eqnarray}
where the boundary is the $d$-dimensional flat space at $z=0$. The relations between the boundary data of AdS and the CFT quantities have been well established in \cite{Witten:1998qj} by the field-operator correspondence. That is, the correlators of a conformal primary operator $\mathcal{O}(x)$ in the CFT are reproduced by the asymptotical data of a bulk field $\phi$ near the boundary.
However, the explicit CFT construction of the bulk local degree of freedoms $\phi(x,z)$ inside the AdS space are not well understood yet. The earlier attempts \cite{Banks:1998dd,Balasubramanian:1999ri,Bena:1999jv} suggested to reconstruct $\phi(x,z)$ by propagating the bulk modes from the bulk to the boundary, and then \cite{Hamilton:2006az} showed that it is equivalent to the smearing operator construction. In this letter, we suggest a different construction based on almost purely CFT arguments. In Section 2.1, we establish the construction by considering CFT at finite energy scale. The possible divergence and the prescription of regulator are discussed in Section 2.2. Then in Section 3, we find that our construction can get improved in the radial quantization formalism. We summarize our main results in Section 4, and a possible way of generalizing the construction beyond pure AdS is also proposed there.
\section{CFT construction of bulk local operators}
\subsection{Renormalized primary at finite energy scale}
It has been  pointed out qualitatively \cite{Maldacena:1997re,Gubser:1998bc,Witten:1998qj,Susskind:1998dq} that the bulk radial direction $z$ is related to the energy scale in the dual field theory. In order to reconstruct $\phi(x,z)$, the first candidate is to consider in CFT the renormalized primary operator $\mathcal{O}(x,\upmu)$ which is defined at a finite energy scale $\upmu$. On the other hand, the behaviors of a primary operator under the conformal transformation have already been encoded in its conformal family. Thus it is natural to expect, at least in the leading order, that the renormalization of the primary operator at a finite energy scale will lead to a mixing between the primary operator and its descendants
\begin{eqnarray}
  {\mathcal{O}}(x,\upmu)=Z(\upmu,\partial){\mathcal{O}}(x)\,.
\end{eqnarray}
For simplicity, we will only consider the scalar operator from now on. It is also natural to require that the renormalized primary operator recovers the Lorentz properties and the scaling dimension of the original primary. Then we find it can only be in the following form
\begin{eqnarray}
  {\mathcal{O}}(x,\upmu)=Z(\upmu^{-2}\Box){\mathcal{O}}(x)\,.
\end{eqnarray}

Is it possible to fix the explicit form of $Z(\upmu^{-2}\Box)$ by imposing certain renormalization condition? The idea is to give the word ``primary'' a renormalized meaning.  In the usual CFT language, the definition of a primary operator is equivalent to require it transforms as a tensor under conformal transformations. We also notice that the renormalization scale $\upmu$ will transform non-trivially under conformal transformations. Thus a direct guess is that the proper renormalization condition should be the follwing:
\\{\bf{The renormalized primary transforms as a tensor under the generalized conformal transformations including the energy scale.}}

To address the generalized conformal transformations including the energy scale, let us firstly review the realization of conformal algebra on the $x$-space. Acting on the coordinates $x^{\mu}$, the conformal generator can be expressed as following
\begin{eqnarray}\label{confx}
P_\mu\circ\!\!\!\!&=&\!\!\!\!-\ii\partial_\mu ,\cr
M_{\mu\nu}\circ\!\!\!\!&=&\!\!\!\!-\ii(x_\mu \partial_\nu - x_\nu \partial_\mu), \cr
D\circ\!\!\!\!&=&\!\!\!\!-\ii x^\mu \partial_\mu,\cr
K_\mu\circ\!\!\!\!&=&\!\!\!\!-\ii(x^2\partial_\mu - 2x_\mu x^\nu \partial_\nu) .
\end{eqnarray}
It implies the standard conformal algebra
\begin{eqnarray}
&&[P_{\mu},P_\nu]=0\,,~~~[M_{\mu\nu},M_{\rho\sigma}]=\ii(\eta_{\mu\rho}M_{\nu\sigma}+\eta_{\nu\sigma}M_{\mu\rho}
-\eta_{\nu\rho}M_{\mu\sigma}-\eta_{\mu\sigma}M_{\nu\rho})\,,
\cr
&&[M_{\mu\nu},P_\rho]=\ii(\eta_{\mu\rho}P_\nu-
\eta_{\nu\rho}P_\mu)\,, ~~~
[M_{\mu\nu},K_\rho]=\ii(\eta_{\mu\rho}K_\nu-\eta_{\nu\rho}K_\mu)\,, ~~~
\cr
&&
[D,M_{\mu\nu}]=0\,, ~~~
[D,K_\mu]=-\ii K_\mu\,, ~~~
[D,P_\mu]=\ii P_\mu\,,~~~
\cr&& [K_{\mu},K_\nu]=0\,,~~~
[K_\mu,P_\nu]=-2\ii M_{\mu\nu}-2\ii\eta_{\mu\nu}D\,.~~~~~~
\end{eqnarray}

To include the energy scale\footnote{For different approaches of introducing the finite energy scale, see \cite{Heemskerk:2010hk,Sarkar:2014jia}.}, a straightforward way is to add $\frac{\partial}{\partial\upmu}$ as well as $\upmu$-dependent coefficients into the realization (\ref{confx}). For latter convenience, we define $z\equiv1/{\upmu}$ and equivalently consider $z$ instead. From the fact that the energy scale is Poincar\'e invariant, we conclude that the forms of $P_{\mu}$ and $M_{\mu\nu}$ remain intact. The scaling dimension of energy scale is obviously 1, thus we can easily write down the following generalized form of dilatation $D$
\begin{eqnarray}
D\circ\!\!\!\!&=&\!\!\!\!-\ii (z\partial_z+x^\mu \partial_\mu)\,.
\end{eqnarray}
For the special conformal generator, the strategy is to take its most general ansatz
\begin{eqnarray}
K_\mu\circ = -\ii\left[x^2\delta_{\mu}^{\nu}- 2x_\mu x^\nu +f_{\mu}{}^{\nu}(z,x)\right]\partial_\nu-\ii g_{\mu}(z,x)\partial_z
\end{eqnarray}
and then try to find the explicit form which satisfies the conformal algebra.
\\From $[K_\mu,P_\nu]=-2\ii M_{\mu\nu}-2\ii\eta_{\mu\nu}D$, we get
\begin{eqnarray}
\!\!\!\!&&\!\!\!\!2\eta_{\mu\nu}z\partial_z= -\partial_{\nu}f_{\mu}{}^{\rho}(z,x)\,\partial_\rho-\partial_{\nu}g_{\mu}(z,x)\,\partial_z\,.
\end{eqnarray}
It implies that
\begin{eqnarray}
f_{\mu}{}^{\nu}(z,x)=\delta_{\mu}^{\nu}f(z)\,,~~~g_{\mu}(z,x)=-2x_{\mu}z\,.
\end{eqnarray}
From $[D,K_\mu]=-\ii K_\mu$, we further get
\begin{eqnarray}
f(z)=\alpha z^2\,,
\end{eqnarray}
where $\alpha$ is an arbitrary constant.
Finally, we can check that the above results satisfy $[K_\mu,K_\nu]=0$. In conclusion, we have
\begin{eqnarray}\label{zconf}
P_\mu\circ\!\!\!\!&=&\!\!\!\!-\ii\partial_\mu \,,\cr
M_{\mu\nu}\circ\!\!\!\!&=&\!\!\!\!-\ii(x_\mu \partial_\nu - x_\nu \partial_\mu)\,,\cr
D\circ\!\!\!\!&=&\!\!\!\!-\ii (z\partial_z+x^\mu \partial_\mu)\,,
\cr K_{\mu}\circ\!\!\!\!&=&\!\!\!\!-\ii\left[( x^2+\alpha z^2)\partial_\mu-2x_{\mu}x^{\rho}\partial_{\rho}-2 x_{\mu} z\partial_{z}\right] .
\end{eqnarray}
In fact, this is exactly the isometry generator of the AdS space when $\alpha>0$ and it suggests to identify $\sqrt{\alpha}z$ here with the standard AdS radial coordinate.
We also notice that it corresponds to dS$_{d+1}$ when $\alpha$ is negative, but we will only concentrate on the AdS case in this paper.

Given the generalized conformal transformation including the energy scale (\ref{zconf}), we can try to decide the form of $Z(z^2\Box)$ by our renormalization condition on primary. For a scalar in the $\{x,z\}$ space, we can expand it by powers of $z$
\begin{eqnarray}
\Phi(z,x)=z^{\Delta}\sum_{n=0}^{\infty}z^n\Phi_n(x)\,.
\end{eqnarray}
The scalar transformation rule
\begin{eqnarray}
  \tilde{\Phi}({\tilde z,\tilde x}) = \Phi(z,x)
\end{eqnarray}
implies that the terms appeared in the power expansion should transform as following
\begin{eqnarray}\label{Bscalar}
[P_\mu, \Phi_n(x)] \!\!\!\!&=&\!\!\!\!\ii\partial_\mu \Phi_n(x)\cr
[M_{\mu\nu}, \Phi_n(x)]\!\!\!\!&=&\!\!\!\!\ii(x_\mu \partial_\nu - x_\nu \partial_\mu) \Phi_n(x), \cr
[D, \Phi_n(x)] \!\!\!\!&=&\!\!\!\!\ii(\Delta+n + x^\mu \partial_\mu) \Phi_n(x),\cr
[K_\mu, \Phi_0(x)] \!\!\!\!&=&\!\!\!\!
 \ii[x^2\partial_\mu - 2x_\mu x^\nu \partial_\nu - 2x_\mu
\Delta] \Phi_0(x),\cr
[K_\mu, \Phi_1(x)] \!\!\!\!&=&\!\!\!\!
 \ii[x^2\partial_\mu - 2x_\mu x^\nu \partial_\nu - 2x_\mu
(\Delta+1)] \Phi_1(x),\cr
[K_\mu, \Phi_n(x)] \!\!\!\!&=&\!\!\!\!
\ii\alpha\partial_\mu\Phi_{n-2}(x)+ \ii[x^2\partial_\mu - 2x_\mu x^\nu \partial_\nu - 2x_\mu
(\Delta+n)] \Phi_n(x)~~(n>1).~
\end{eqnarray}
Now the task is to construct $\Phi_n(x)$ by the primary $\mathcal{O}$ and it scalar descendents $\Box^m\mathcal{O}$.
From the conformal transformation rules of the primary
we can deduce that
\begin{eqnarray}
[P_\mu, \Box^n\mathcal{O}(x)]\!\!\!\!&=&\!\!\!\!\ii\partial_\mu \Box^n\mathcal{O}(x)\,,~\cr
[M_{\mu\nu}, \Box^n\mathcal{O}(x)]\!\!\!\!&=&\!\!\!\!\left[\ii(x_\mu \partial_\nu - x_\nu \partial_\mu)+\Sigma_{\nu\rho}^{(\mathcal{O})}\right] \Box^n\mathcal{O}(x)\,,\cr
[D, \Box^n\mathcal{O}(x)]\!\!\!\!&=&\!\!\!\!\ii(\Delta+2n + x^\mu \partial_\mu) \Box^n\mathcal{O}(x)\,,\cr
[K_\mu, \mathcal{O}(x)]\!\!\!\!&=&\!\!\!\![\ii(x^2\partial_\mu - 2x_\mu x^\nu \partial_\nu - 2x_\mu
\Delta) - 2x^\nu \Sigma_{\mu\nu}^{(\mathcal{O})} ] \mathcal{O}(x)\,,\cr
[K_{\mu},\Box^n \mathcal{O}(x)]
\!\!\!\!&=&\!\!\!\!2n\left[d-2(\Delta+n)\right]\ii\,\partial_{\mu}\Box^{n-1}\mathcal{O}(x)
-4n\Sigma_{\mu\nu}^{(\mathcal{O})}\partial^{\nu}\Box^{n-1}\mathcal{O}(x)~~~~~~
\cr&&+\left[\ii(x^2\partial_{\mu}-2x_{\mu}x^{\rho}\partial_{\rho}-2x_{\mu}(\Delta+2n))
-2x^{\rho}\Sigma_{\mu\rho}^{(\mathcal{O})}\right]\Box^{n}\mathcal{O}(x)~~(n\geq1)\,.~~~~~~
\end{eqnarray}
Comparing with (\ref{Bscalar}), it implies the unique identification
\begin{eqnarray}
\Phi_{2n+1}=0\,,~~~
\Phi_{2n}=\frac{(-1)^n\,\alpha^{n}\,\Gamma(\Delta-\tfrac{d}2+1)} {4^n\,n!\,\Gamma\!\left(\Delta-\tfrac{d}2+n+1\right)}\Box^n{\mathcal{O}}\,.
\end{eqnarray}
In conclusion, up to an overall constant, our arguments show that the renormalized primary at energy scale $\upmu=1/z$ is given by
\begin{eqnarray}\label{map}
\mathcal{O}(x,z)=Z(z^2\Box)\mathcal{O}(x)={}_0F_1\!\left(;\Delta-\tfrac{d}2+1;-\tfrac{\alpha z^2}{4}\Box\right)\mathcal{O}(x)\,,
\end{eqnarray}
and $z^{\Delta}\mathcal{O}(x,z)$ corresponds to a bulk scalar field $\phi(x,z)$. In the $\alpha=0$ limit, it actually comes back to the usual language of CFT.

We notice that $Z(z^2\Box)$ obtained in (\ref{map}) is nothing but the Fourier transformation of the solutions to the bulk free field equation  with $\Phi(z\to0)\sim z^{\Delta}$ behavior at the boundary. This construction is different from the one suggested in \cite{Banks:1998dd,Balasubramanian:1999ri,Bena:1999jv,Hamilton:2006az}. The approach there encountered only the $k^2={\vec {\rm k}}^2-w^2<0$ part of the bulk $\Phi(z\to0)\sim z^{\Delta}$ modes, and thus it can not be generalized to the Euclidean AdS case. Instead, our construction encounters all the bulk $\Phi(z\to0)\sim z^{\Delta}$ modes since it is the honest Fourier transformation. Obviously, (\ref{map}) is applicable for both signatures.

\subsection{Two point correlators: the divergent regime and the regulator}
As a consistency check, let us use (\ref{map}) to recover the well-known bulk-boundary propagator. We find
\begin{eqnarray}\label{bB1}
z^{\Delta}\left\langle\mathcal{O}(z,x)\mathcal{O}(x')\right\rangle\!\!\!\!&=&\!\!\!\!
   \left\langle{z^{\Delta}}{}_0F_1\!\left(;\Delta-\tfrac{d}2+1;-\tfrac{\alpha z^2}{4}\Box\right)\mathcal{O}(x)\,\mathcal{O}(x')\right\rangle
\cr\!\!\!\!&=&\!\!\!\!\frac{z^{\Delta}}{(x-x')^{2\Delta}}\sum_{n=0}^{\infty}
   \frac{\Delta(\Delta+1)\cdots(\Delta+n-1)}{n!}\left(\frac{-\alpha z^{2}}{(x-x')^{2}}\right)^n\,.
\end{eqnarray}
In the regime $|x-x'|^2>\alpha z^2$, the series is convergent and gives rise to the expected form of the bulk-boundary propagator
\begin{eqnarray}\label{bB}
K(x,z;x')\!\!\!\!&=&\!\!\!\!\left(\frac{z}{\alpha z^2+(x-x')^2}\right)^{\Delta}\,.
\end{eqnarray}
However, in the regime $|x-x'|^2<\alpha z^2$, the series (\ref{bB1}) is divergent. In fact, this result is not surprising. The $Z(z^2\Box)$ given in (\ref{map}) is just the $\Phi(z\to0)\sim z^{\Delta}$ modes of the bulk solution,
while the Fourier transformation of (\ref{bB}) is a linear combination of the $\Phi(z\to0)\sim z^{\Delta}$ modes and the $\Phi(z\to0)\sim z^{d-\Delta}$ modes \cite{Balasubramanian:1998sn} which regulate the divergence of (\ref{bB1}).
The existence of the $\Phi(z\to0)\sim z^{d-\Delta}$ constituent in (\ref{bB}) can be easily seen from the $z\to0$ limit \cite{Freedman:1998tz}
\begin{eqnarray}
\lim_{z\to 0}\left(\frac{z}{\alpha z^2+(x-x')^2}\right)^{\Delta}\sim z^{d-\Delta}\delta(x-x')\,.
\nn
\end{eqnarray}
Although both the $\Phi(z\to0)\sim z^{\Delta}$ and $\Phi(z\to0)\sim z^{d-\Delta}$ modes diverge exponentially
as $z\to\infty$, the combination is well-behaved in the interior the interior since the two divergences cancel with each other. The explicit computation of the corresponding Fourier transformations are performed in the Appendix.

In order to understand the above issue better, let us recall a simple fact in field theory. That is, the correlation function for composite operators always have zero-th order UV divergence due to its composite natural. For example, consider the composite operator 
${\mathcal{O}}=:\phi_a\phi^a:$.
The two point correlator $\langle{\mathcal{O}}{\mathcal{O}}\rangle$  receives zero-th order UV divergence from the following loop diagram even in the free theory.
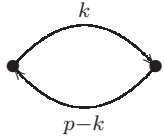
\begin{equation}
\begin{xy} 
 \xymatrix{
 *=0{\bullet} \ar^{k}@/^1.3pc/[rr]
 &
 & *=0{\bullet} \ar^{p-k}@/^1.3pc/[ll]
 }
\end{xy}\nn
\end{equation}
In the coordinate space, the corresponding divergence takes the following form
\begin{equation}
f(\Lambda,\upmu,\Box)\delta^d(x-x')
\end{equation}
where $\upmu$ is the renormalization scale and $\Lambda$ is the cut off scale.
This divergence can not be canceled by any local counterterm in the original action. Instead, we need to define the regularized two point function directly and remove it by hand. Or equivalently speaking, we need to add a local counterterm in the free energy $W[J]$
\begin{eqnarray}
e^{W[J]}=\left\langle e^{J\mathcal{O}}\right\rangle\,,~~~
W[J]\rightarrow W_{\rm R}[J]=W[J]-\int\rd^d x\int\rd^d x'\, c(x-x')J(x)J(x')\,,\nn
\end{eqnarray}
where $c(x-x')=c(\Lambda,\upmu,\Box)\delta^d(x-x')$.
In principle, after the cancellation of the divergent part, a possible remnant term in the form
$R(\upmu,\Box)\delta^d(x-x')$ would still be there, and its explicit form depends on the prescription of the regularization.

We notice that the $\Phi(z\to0)\sim z^{d-\Delta}$ modes of (\ref{bB}) in the coordinate space is given by
\begin{eqnarray}\label{non}
z^{d-\Delta}{}_0F_1\!\left(;\frac{d}{2}-\Delta+1;-\frac{\alpha z^2}{4}\Box\right)
\delta^d(x-x')\,.
\end{eqnarray}
It is right in the form of $R(\upmu,\Box)\delta^d(x-x')$ appeared above. This fact suggests that one can understand it as the possible remnant term. The only special point is that there is a infinite order derivative operator acting on $\delta^d(x-x')$. Thus it is no longer a local function but a quasi-local term which is identically vanishing in the outer region $|x-x'|^2>\alpha z^2$. Adding such a term does not affect the result (\ref{bB1}) in the region $|x-x'|^2>\alpha z^2$, and is possible to cancel the divergence in the region $|x-x'|^2<\alpha z^2$. If we take the continuity at $|x-x'|^2=\alpha z^2$ as the prescription of the regularization of the two point function, it will pick the correct ratio between the $\Phi(z\to0)\sim z^{\Delta}$ modes and the $\Phi(z\to0)\sim z^{d-\Delta}$ modes, then recovers (\ref{bB}) everywhere. This prescription is equivalent to the momentum space IR regularity condition used in the literatures \cite{Balasubramanian:1998sn}. Since it is natural to expect that effective operators defined at finite energy scale have some ambiguity in probing the distance shorter than its typical scale, the dependence on the prescription of regularization above are actually acceptable.

One can also check that the bulk-bulk propagator can be recovered by computing
\begin{eqnarray}
z^{\Delta}z'^{\Delta}\left\langle\mathcal{O}(z,x)\mathcal{O}(z',x')\right\rangle\,.
\end{eqnarray}
Again, there is a divergent regime at short distance. If the continuity prescription is imposed, it implies that one should take the following regulator
\begin{eqnarray}\label{non}
z_M^{d-\Delta}z_m^{\Delta}\,{}_0F_1\!\left(;\frac{d}{2}-\Delta+1;-\frac{\alpha z_M^2}{4}\Box\right){}_0F_1\!\left(;\Delta-\frac{d}{2}+1;-\frac{\alpha z_m^2}{4}\Box\right)
\delta^d(x-x')
\end{eqnarray}
where $z_M=max\{z,z'\}$ and $z_m=min\{z,z'\}$.

\section{Radial quantization}
\subsection{Radial quantization in CFT}
In Section 2, we have constructed the bulk local operator and also explained its divergent regime with the regularization prescription there. However, the present formula is not convenient in discussing the bulk physics since the regulator should always be added by hand. It will be pretty nice if one can find a smart formula in which the regulator has been automatically built in. To achieve such a formula, let us discuss the radial quantization in usual CFT language firstly.

The radial quantization for CFT was detailedly reviewed in \cite{Pappadopulo:2012jk}. We will equivalently re-express the results there by introducing the radial expansion for the operators. Again, we will just consider scalar primary here.
The radial expansion of a scalar primary operator ${\mathcal{O}}(x)$ is given by
\begin{eqnarray}\label{Rexp}
{\mathcal{O}}(x)\!\!\!\!&=&\!\!\!\!\sum_{m=0}^{\infty}\frac{1}{m!}x^{\mu_1}\cdots x^{\mu_m}O_{\mu_1\cdots\mu_m}
+\sum_{n=0}^{\infty}\frac{1}{n!}\check x^{2\Delta}\check x^{\nu_1}\cdots \check  x^{\nu_n}\check O_{\nu_1\cdots\nu_n}\,,
\end{eqnarray}
where $\check x$ is the inversion of $x$
\begin{eqnarray}
\check x^{\mu}=\mathcal{I}\circ x^{\mu}=\frac{x^{\mu}}{x^2}\,.
\end{eqnarray}
In an unitary theory, its Hermitian conjugation is induced by the inversion
\begin{eqnarray}
{\mathcal{O}}^{\dagger}(x)\!\!\!\!&=&\!\!\!\!\check{\mathcal{O}}(-x)=\frac{1}{x^{2\Delta}}{\mathcal{O}}(-\check x)
\cr\!\!\!\!&=&\!\!\!\!\sum_{n=0}^{\infty}\frac{(-1)^n}{n!}x^{\nu_1}\cdots x^{\nu_n}\check O_{\nu_1\cdots\nu_n}
+\sum_{m=0}^{\infty}\frac{(-1)^m}{m!}\check x^{2\Delta}\check x^{\mu_1}\cdots \check  x^{\mu_m}O_{\mu_1\cdots\mu_m}\,.
\end{eqnarray}
In terms of the component operators, it is given by
\begin{eqnarray}\label{Hconj}
O^{\dagger}_{\mu_1\cdots\mu_n}=(-1)^n\check O_{\mu_1\cdots\mu_n}\,,~~~\check O^{\dagger}_{\mu_1\cdots\mu_n}=(-1)^nO_{\mu_1\cdots\mu_n}\,.
\end{eqnarray}
The vacuum $|1\rangle$ is defined by
\begin{eqnarray}\label{vac}
\check O_{\nu_1\cdots\nu_n}|1\rangle=0\,,~~~~~~~~~~
\langle1|O_{\mu_1\cdots\mu_m}=0\,.
\end{eqnarray}
It is equivalent to requiring that the state ${\mathcal{O}}(x)|1\rangle$ and all its descendants are regular at $x=0$, while the conjugated state
$\langle1|\check{\mathcal{O}}(x)$ and all its descendants are also regular at $x=0$.
One can check that the vacuum $|1\rangle$ defined in (\ref{vac}) is actually conformal invariant. 
The conformal transformation rules of the component operators can be deduced from the standard rules for primary ${\mathcal{O}}(x)$. We find
\begin{eqnarray}\label{confRa}
\!\!\!\!&&\!\!\!
[P_{\mu},O_{\mu_1\dots\mu_m}]=\ii O_{\mu\mu_1\cdots\mu_m}\,,~~~
\cr\!\!\!\!&&\!\!\!
[P_{\mu},\check O_{\nu_1\dots\nu_n}]=\ii\left[n(n-1)\eta_{(\nu_1\nu_2}\check O_{\nu_3\cdots\nu_n)\mu}-2n(\Delta+n-1)\eta_{\mu(\nu_1}\check O_{\nu_2\cdots\nu_n)}\right]\,,~~~~
\cr\!\!\!\!&&\!\!\!
[M_{\mu\nu}, O_{\mu_1\dots\mu_m}]=\ii m\left[\eta_{\mu(\mu_1} O_{\mu_2\cdots\mu_m)\nu}-\eta_{\nu(\mu_1} O_{\mu_2\cdots\mu_m)\mu}\right]\,,~~~
\cr\!\!\!\!&&\!\!\!
[M_{\mu\nu},\check O_{\nu_1\dots\nu_n}]=\ii n\left[\eta_{\mu(\nu_1} \check O_{\nu_2\cdots\nu_n)\nu}-\eta_{\nu(\nu_1} \check O_{\nu_2\cdots\nu_m)\mu}\right]\,,~~~
\cr\!\!\!\!&&\!\!\!
[D,O_{\mu_1\dots\mu_m}]=\ii (\Delta+m)O_{\mu_1\dots\mu_m}\,,~~~
\cr\!\!\!\!&&\!\!\!
[D,\check O_{\nu_1\dots\nu_n}]=-\ii(\Delta+m)\check O_{\nu_1\cdots\nu_n}\,,~~~~
\cr\!\!\!\!&&\!\!\!
[K_{\mu},O_{\mu_1\dots\mu_m}]=\ii\left[ m(m-1)\eta_{(\mu_1\mu_2}O_{\mu_3\cdots\mu_m)\mu}
-2 m(\Delta+m-1)\eta_{\mu(\mu_1}O_{\mu_2\cdots\mu_m)}\right]\,,~~~
\cr\!\!\!\!&&\!\!\!
[K_{\mu},\check O_{\nu_1\dots\nu_n}]=\ii \check O_{\mu\nu_1\cdots\nu_n}\,.
\end{eqnarray}
Given the input data
\begin{eqnarray}
\langle{\mathcal{O}}'(0)|{\mathcal{O}}(0)\rangle=\langle1|\check O_{0}'O_{0}|1\rangle=C_{\mathcal{O}'\mathcal{O}}\,,
\end{eqnarray}
one can decide the inner product between the states $|\partial_{\mu_n}\cdots\partial_{\mu_1}\mathcal{O}(0)\rangle
=O_{\mu_1\dots\mu_n}|1\rangle$ by using 
(\ref{confRa}) and the conformal invariance of the vacuum. The result is
\begin{eqnarray}
\!\!\!\!&&\!\!\!\!\langle1|\check O_{\nu_n\cdots\nu_1}'O^{\mu_1\cdots\mu_m}|1\rangle
\cr\!\!\!\!&=&\!\!\!\!\delta_{mn}C_{\mathcal{O'O}}\sum_{k=0}^{\lfloor\frac{n}2\rfloor}\frac{(-1)^{n-k}\,2^n\,(n!)^2\,\Gamma(\Delta+n-k)}{(n-2k)!\,k!\,(2!)^{2k}\,\Gamma(\Delta)}
\cr&&~~~~~~~~~~~~~~~\delta^{(\mu_1}_{(\nu_1}\cdots\delta^{\mu_{n-2k}}_{\nu_{n-2k}}
\eta_{\nu_{n-2k+1}\nu_{n-2k+2}}\cdots\eta_{\nu_{n-1}\nu_{n})}\eta^{\mu_{n-2k+1}\mu_{n-2k+2}}\cdots\eta^{\mu_{n-1}\mu_{n})}\,.~~~~~~
\end{eqnarray}
Now we can reproduce the well-known two point  correlator by the ``silly'' computation
\begin{eqnarray}\label{2ptR}
\langle1|{\mathcal{O}}(x){\mathcal{O}}'(x')|1\rangle
\!\!\!\!&=&\!\!\!\!\sum_{m=0}^{\infty}\sum_{n=0}^{\infty}\frac{1}{m!\,n!}\frac{x^{\mu_1}\cdots x^{\mu_m}x'^{\nu_1}\cdots x'^{\nu_n}}{x^{2(\Delta+m)}}
\langle1|\check O_{\mu_1\cdots\mu_m}\,O'_{\nu_1\cdots\nu_n}|1\rangle
\cr\!\!\!\!&=&\!\!\!\!C_{\mathcal{O'O}}\sum_{\hat n=0}^{\infty}\sum_{k=0}^{\infty}
\frac{(-1)^{\hat n+k}\,2^{\hat n}\,\Gamma(\Delta+\hat n+k)}{\hat n!\,k!\,\Gamma(\Delta)}
\frac{(x\cdot x')^{\hat n}x'^{2k}}{x^{2(\Delta+\hat n+k)}}\,.
\end{eqnarray}
The convergence of the series requires that $|x|>|x'|.${\footnote{Strictly speaking, the convergence argument is accurate only for the Euclidean case. For the Lorentzian case, proper analytical continuations are needed as what usually happened in the quantum field theory computations.}} It means that the usual CFT correlator is reproduced by the radial ordered function $\langle1|\hat{\mathcal{R}}\,{\mathcal{O}}(x){\mathcal{O}}'(x')|1\rangle$. In the radial quantization where the dilatation operator $D$ is treated as the Hamiltonian, the radial ordered function is the natural analogy of the time ordered function in the usual quantum field theory.

\subsection{Radial quantization at finite energy scale}
Now let us consider the CFT radial quantization in the presence of the finite energy scale $\upmu$. A direct idea is acting the $Z(\upmu^{-2}\Box)$ in (\ref{map}) onto the radial expansion (\ref{Rexp}). We get
\begin{eqnarray}
\!\!\!\!&&\!\!\!\!
{}_0F_1\!\left(;\Delta-\tfrac{d}2+1;-\tfrac{\alpha z^2}{4}\Box\right)\mathcal{O}(x)
\cr\!\!\!\!&=&\!\!\!\!\sum_{s=0}^{\infty}\sum_{k=0}^{\infty}\frac{(-\alpha z^2)^k\,\Gamma\!\left(\Delta-\tfrac{d}2+1\right)}
{2^{2k}\,k!\,s!\,\Gamma\!\left(\Delta-\tfrac{d}2+k+1\right)}
x^{\nu_{1}}\cdots x^{\nu_s}O_{\mu_1\cdots\mu_k}{}^{\mu_1\cdots\mu_k}{}_{\nu_{1}\cdots\nu_s}
\cr\!\!\!\!&&\!\!\!\!+\sum_{s=0}^{\infty}\sum_{l=0}^{\infty}\frac{(-\alpha z^2)^{l}\,\Gamma\!\left(\Delta-\tfrac{d}2+1\right)}
{2^{2l}\,l!\,s!\,\Gamma\!\left(\Delta-\frac{d}2+l+1\right)}{}_1F_0\!\left(\Delta+2l+s;;-\tfrac{\alpha z^2}{x^2}\right)
\frac{x^{\nu_{1}}\cdots x^{\nu_s}}{x^{2(\Delta+2l+s)}}\check O_{\mu_1\cdots\mu_l}{}^{\mu_1\cdots\mu_l}{}_{\nu_{1}\cdots\nu_s}
\,.~~~~~~
\end{eqnarray}
We notice that when $|x|^2>\alpha z^2$
\begin{eqnarray}
\frac{1}{x^{2(\Delta+2l+s)}}{}_1F_0\!\left(\Delta+2l+s;;-\tfrac{\alpha z^2}{x^2}\right)=\frac{1}{(x^2+\alpha z^2)^{\Delta+2l+s}}\,,
\end{eqnarray}
and it is divergent in the regime $|x|^2<\alpha z^2$. The structure of the divergent regime is quite similar to what we have seen in Section 2.
Thus it is natural to expect the following as the regularized radial expansion at finite energy scale
\begin{eqnarray}\label{Rexpz}
{\mathcal{O}}(x,z)\!\!\!\!&=&\!\!\!\!\sum_{m=0}^{\infty}\frac{1}{m!}x^{\mu_1}\cdots x^{\mu_m}O_{\mu_1\cdots\mu_m}(z) +\sum_{n=0}^{\infty}\frac{1}{n!}
\frac{x^{\nu_1}\cdots x^{\nu_n}}
{(x^2+\alpha z^2)^{\Delta+n}}\check O_{\nu_1\cdots\nu_n}(\check z)
\end{eqnarray}
where the component operator at finite energy scale are linear combinations of the original ones
\begin{eqnarray}
O_{\mu_1\cdots\mu_m}(z)\!\!\!\!&=&\!\!\!\!\sum_{k=0}^{\infty}\frac{(-\alpha z^2)^k\,\Gamma\!\left(\Delta-\tfrac{d}2+1\right)}
{2^{2k}\,k!\,\Gamma\!\left(\Delta-\tfrac{d}2+k+1\right)}O_{\nu_1\cdots\nu_k}{}^{\nu_1\cdots\nu_k}{}_{\mu_{1}\cdots\mu_m}\,,
\cr\check O_{\mu_1\cdots\mu_m}(\check z)\!\!\!\!&=&\!\!\!\!\sum_{k=0}^{\infty}\frac{(-\alpha \check z^2)^k\,\Gamma\!\left(\Delta-\tfrac{d}2+1\right)}
{2^{2k}\,k!\,\Gamma\!\left(\Delta-\tfrac{d}2+k+1\right)}\check O_{\nu_1\cdots\nu_k}{}^{\nu_1\cdots\nu_k}{}_{\mu_{1}\cdots\mu_m}\,,
\end{eqnarray}
and the inversion of $x$ and $z$ in the generalized scene are given by
\begin{eqnarray}
{\check x}^\mu=\mathcal{I}\circ x^{\mu}= {{x^\mu } \over {x^2+\alpha z^2}}\,,~~~~~~
{\check z}=\mathcal{I}\circ z={z \over {x^2+\alpha z^2}}\,.
\end{eqnarray}
Correspondingly, the inversion of ${\mathcal{O}}(x,z)$ is
\begin{eqnarray}
\check{\mathcal{O}}(x,z)\!\!\!\!&=&\!\!\!\!\frac{1}{(x^2+\alpha z^2)^{\Delta}}{\mathcal{O}}(\check x;\check z)
\cr\!\!\!\!&=&\!\!\!\!\sum_{n=0}^{\infty}\frac{1}{n!}x^{\nu_1}\cdots x^{\nu_n}\check O_{\nu_1\cdots\nu_n}(z) +\sum_{m=0}^{\infty}\frac{1}{m!}\frac{x^{\mu_1}\cdots x^{\mu_m}}{(x^2+\alpha z^2)^{\Delta+n}}O_{\mu_1\cdots\mu_m}(\check z)\,.
\end{eqnarray}
From (\ref{Hconj}), we can see that the Hermitian conjugation relation keeps intact at finite energy scale
\begin{eqnarray}
{\mathcal{O}}^{\dagger}(x,z)\!\!\!\!&=&\!\!\!\!\check{\mathcal{O}}(-x,z)\,.
\end{eqnarray}

Simply by using (\ref{confRa}), one can write down the conformal transformation rule for the component fields at finite energy scale directly as following
\begin{eqnarray}\label{CTransz}
\!\!\!\!&&\!\!\!\![P_{\mu},O_{\mu_1\cdots\mu_m}(z)]
=\ii\,O_{\mu\mu_1\cdots\mu_m}(z)\,,
\cr
\!\!\!\!&&\!\!\!\!
[P_{\mu},\check O_{\mu_1\cdots\mu_m}(z)]=\ii
\Big[\alpha z^2\check O_{\mu_{1}\cdots\mu_m\mu}(z)
+m(m-1)\eta_{(\mu_1\mu_2}\check O_{\mu_{3}\cdots\mu_m)\mu}(z)
\cr&&~~~~~~~~~~~~~~~~~~~~~~~~~~
-2m(\Delta+m-1+z\partial_z)\eta_{\mu(\mu_1}\check O_{\mu_{2}\cdots\mu_m)}(z)\Big]\,,
\cr
\!\!\!\!&&\!\!\!
[M_{\mu\nu}, O_{\mu_1\dots\mu_m}(z)]=\ii m\left[\eta_{\mu(\mu_1} O_{\mu_2\cdots\mu_m)\nu}(z)-\eta_{\nu(\mu_1} O_{\mu_2\cdots\mu_m)\mu}(z)\right]\,,
\cr
\!\!\!\!&&\!\!\!
[M_{\mu\nu},\check O_{\mu_1\dots\mu_m}(z)]=\ii m\left[\eta_{\mu(\nu_1} \check O_{\nu_2\cdots\nu_m)\nu}(z)-\eta_{\nu(\nu_1} \check O_{\nu_2\cdots\nu_m)\mu}(z)\right]\,,
\cr
\!\!\!\!&&\!\!\!
[D,O_{\mu_1\dots\mu_m}(z)]=\ii(\Delta+m+z\partial_z)O_{\mu_1\dots\mu_m}(z)\,,
\cr
\!\!\!\!&&\!\!\!
[D,\check O_{\mu_1\dots\mu_m}(z)]=-\ii(\Delta+m+z\partial_z)\check O_{\mu_1\dots\mu_m}(z)\,,
\cr
\!\!\!\!&&\!\!\!
[K_{\mu},O_{\mu_1\dots\mu_m}(z)]=\ii
\Big[\alpha z^2  O_{\mu_{1}\cdots\mu_m\mu}(z)
+m(m-1)\eta_{(\mu_1\mu_2}  O_{\mu_{3}\cdots\mu_m)\mu}(z)
\cr&&~~~~~~~~~~~~~~~~~~~~~~~~~~~~-2m(\Delta+m-1+z\partial_z)\eta_{\mu(\mu_1}  O_{\mu_{2}\cdots\mu_m)}(z)\Big]\,,
\cr
\!\!\!\!&&\!\!\!
[K_{\mu},\check O_{\mu_1\dots\mu_m}(z)]
=\ii\,\check O_{\mu\mu_1\dots\mu_m}(z)\,.~~~
\end{eqnarray}
Similar to the arguments in Section 2.1, we can show that a bulk scalar field $\Phi(x,z)$ expanded as
\begin{eqnarray}
\Phi(x,z)=\sum_{m=0}^{\infty}\frac{z^{\Delta}}{m!}x^{\mu_1}\cdots x^{\mu_m}\Phi_{\mu_1\cdots\mu_m}(z) +\sum_{n=0}^{\infty}\frac{1}{n!}
\frac{z^{\Delta}\,x^{\nu_1}\cdots x^{\nu_n}}
{(x^2+\alpha z^2)^{\Delta+n}}\check \Phi_{\nu_1\cdots\nu_n}(\check z)
\end{eqnarray}
indeed requires the transformation rules (\ref{CTransz}) for its components.
Thus, the regularized version (\ref{Rexpz}) of  primary at finite energy scale satisfies our basic renormalization condition for primary operators.

Parallel with (\ref{2ptR}), the standard bulk-boundary as well as bulk-bulk propagator can be reproduced by computing the radial ordered function $\langle1|\hat{\mathcal{R}}\,{\mathcal{O}}(x,z){\mathcal{O}}'(x',z')|1\rangle$. In our present case where energy scales are introduced, the radial order is defined by $x^2+\alpha z^2$. Providing this radial order, there is no ambiguity everywhere in reproducing the bulk-boundary and bulk-bulk propagator. Therefore, (\ref{Rexpz}) is
indeed the smart formula which we are looking for.

Since our construction directly comes back to the standard CFT language in the $\alpha\to0$ limit, it is possible that (\ref{Rexpz}) will not suffer from the problem about bulk locality appeared in the smearing operator construction \cite{Kabat:2011rz}. In order to address it properly, one should generalize the standard results about OPE to the cases with finite energy scales. We hope to report on this issue in a future work.

The formula (\ref{Rexpz}) also suggests that one could define the finite energy scale effective Hilbert space $\mathcal{H}_z$ by acting $O_{\mu_1\cdots\mu_m}(z)$ on the vacuum $|1\rangle$. An interesting observation is that $\mathcal{H}_z$ actually contains less information than the UV Hilbert space $\mathcal{H}=\mathcal{H}_{z=0}$.
For example, considering the scalar sector, one can construct the following state
\begin{eqnarray}
|\lambda\rangle=\sum_{n=0}^{\infty}\frac{\lambda^n\,\Gamma(\Delta)\,\Gamma\!\left(\Delta-\tfrac{d}2+1\right)\,\Gamma\!\left(\tfrac{d}2\right)}
{2^{4n}\,n!\,\Gamma(\Delta+n)\,\Gamma\!\left(\Delta-\tfrac{d}2+n+1\right)\,\Gamma\!\left(n+\tfrac{d}2\right)} O_{n}|1\rangle
\,,
\end{eqnarray}
where we denote $O_{n}(z)=O_{\nu_1\cdots\nu_n}{}^{\nu_1\cdots\nu_n}(z)$. It is a well defined state in $\mathcal{H}$ since the norm is finite
\begin{eqnarray}
\langle\lambda|\lambda\rangle=C_{\mathcal{O}}\sum_{n=0}^{\infty}\frac{\lambda^n\,\Gamma(\Delta)\,\Gamma\!\left(\Delta-\tfrac{d}2+1\right)\,\Gamma\!\left(\tfrac{d}2\right)}
{2^{4n}\,n!\,\Gamma(\Delta+n)\,\Gamma\!\left(\Delta-\tfrac{d}2+n+1\right)\,\Gamma\!\left(n+\tfrac{d}2\right)}
\,.
\end{eqnarray}
The inner product between the state $|\lambda\rangle$ and the states in $\mathcal{H}_z$ are given by
\begin{eqnarray}
\langle\lambda|O_{m}(z)|1\rangle=C_{\mathcal{O}}\lambda^m\sum_{k=0}^{\infty}\frac{(-\alpha \lambda z^2)^k\,\Gamma\!\left(\Delta-\tfrac{d}2+1\right)}
{2^{2k}\,k!\,\Gamma\!\left(\Delta-\tfrac{d}2+k+1\right)}
\,.
\end{eqnarray}
Since the ${}_0F_1$ function has infinite number of zeros, the inner product will be zero for infinite many $\lambda$'s providing $z\neq0$.
Therefore, there are infinite number of states in $\mathcal{H}$ perpendicular to the finite energy scale effective Hilbert space $\mathcal{H}_{z\neq0}$.
Although this observation is something one could expect for effective descriptions at finite energy scale, it may has some possible advantages in discussing the $c$-theorem and the entanglement entropy.
\section{Discussions}
In the previous sections, we suggest a CFT construction of the bulk local operators in pure AdS space. The construction is base on considering CFT at finite energy scale.
The basic result is that bulk operator is given by the acting an infinite order different operator $Z_{\rm AdS}(\upmu,\partial)$ onto the original CFT primary.
Although we do not assume any knowledge about the bulk in advance, our arguments automatically show that $Z_{\rm AdS}(\upmu,\partial)$ should be the Fourier transformation of the $\Phi(z\to0)\sim z^{\Delta}$ solution to the bulk free field equation. We also discuss the relation between the regulator of the two point function and the $\Phi(z\to0)\sim z^{d-\Delta}$ modes. In Section 3, based on the radial quantization in CFT, we find an improved formula of our construction in which the regulator is automatically built in.

The next challenge is how to generalize our construction to geometries beyond pure AdS. A naive guess is that the bulk local operator is also effectively given by acting the infinite order different operator $Z_{\rm Geom}(\upmu,\partial)$, which is the Fourier transformation of the bulk $\Phi(z\to0)\sim z^{\Delta}$ modes in the corresponding geometry, onto the original CFT primary. On the other hand, our CFT arguments in Section 2.1 seems state independent. Thus, it suggests that the bulk local operator should always be given by $Z_{\rm AdS}(\upmu,\partial)$ for all asymptotic AdS geometries which are basically very heavy excited states in the CFT. We conjecture that this two possibilities are actually complementary
to each other. The explicit proposal \cite{zlwi013} for the underlying mechanism can be summarized as the following:

Bulk geometries are actually dual to the coherent states $|{\rm Geom}\rangle=F(T_{\mu\nu})|1\rangle$ which is created by acting certain function $F(T_{\mu\nu})$ of stress tensor and its descendants on the vacuum $|1\rangle$.
The bulk correlators of the dual field $\phi(x,z)$ can be reproduced by computing
$$\langle{\rm Geom}|\cdots Z_{\rm AdS}(\upmu,\partial)\mathcal{O}(x)\cdots|{\rm Geom}\rangle
=\langle1|F^{\dagger}(T_{\mu\nu})\cdots Z_{\rm AdS}(\upmu,\partial)\mathcal{O}(x)\cdots F(T_{\mu\nu})|1\rangle\,.$$
On the other hand, by using the local conformal Ward identity\cite{Belavin:1984vu,Fradkin:1996is,Fradkin:1997df}, one may convert (at least in the two dimensional CFT)
the effects of $F(T_{\mu\nu})$ to a differential operator  $\hat{\mathcal{F}}[F]$ acting on the operator $Z_{\rm AdS}(\upmu,\partial)\mathcal{O}(x)$ as following
$$\langle1|F^{\dagger}(T_{\mu\nu})\cdots Z_{\rm AdS}(\upmu,\partial)\mathcal{O}(x)\cdots F(T_{\mu\nu})|1\rangle
=\langle1|\cdots \hat{\mathcal{F}}[F]Z_{\rm AdS}(\upmu,\partial)\mathcal{O}(x)\cdots |1\rangle\,.$$
The new differential operator $\hat{\mathcal{F}}[F]Z_{\rm AdS}(\upmu,\partial)$ is expected to be exactly the $Z_{\rm Geom}(\upmu,\partial)$ of the corresponding geometry. For the black hole geometry, the horizon is the position where the series in $Z_{\rm Geom}(\upmu,\partial)$ becomes ill-defined. However, everything could be still well defined after coming back to the $|{\rm Geom}\rangle$ description and one can explore the black hole interior in this formalism.

Finally, it is also possible that multiple states with different $F(T_{\mu\nu})$ give rise to same differential operator $Z_{\rm Geom}(\upmu,\partial)$. Thus it could be a dual CFT way to explain the entropy of AdS black hole. If indeed so, it means that all the black hole microstates should correspond exactly to the same geometry, and thus one do not need to take any average over different micro-geometries. This picture seems different from what people usually expected for quantum gravity, and may offer new possibilities to the discussions of the black hole firewall problem \cite{Almheiri:2012rt,Braunstein:2009my}.

\vskip 5mm

As the early version of this work was drawing to conclusion, Ref.\cite{Nakayama:2015mva} appeared with results which partially overlap with Section 2.1 in this manuscript. We also realized that the two dimensional version of their results have already appeared in Ref.\cite{Miyaji:2015fia}.

\vskip 5mm

\vskip 1cm
\centerline{\bf\large Acknowledgments}
\vskip 5mm\noindent
The authors are grateful to Piljin Yi for enlightening discussions as well as collaboration at an early stage
of this work, and also thanks Jiang Long, Hong L\"u, Jun-Bao Wu, Xiao Xiao and Hossein Yavartanoo for useful conversations.
This work is supported by National Natural Science Foundation of China with grant No.11305125, No.11447607 and the Double First-class University Construction Project of Northwest University. The authors declare that there is no conflict of interest regarding the publication of this paper.

\appendix
\section{The Momentum space formula}
In the momentum space, the  general solutions to the bulk free scalar equation is given by the linear combination of the $\Phi(z\to0)\sim z^{\Delta}$ modes
\begin{eqnarray}
z^{\Delta}k^{2\Delta-d}\;_0F_1\!\left(;\Delta-\tfrac{d}2+1; \tfrac{z^2k^2}{4}\right)
=\Gamma\!\left(\Delta-\tfrac{d}2+1\right)z^{\tfrac{d}2}(2k)^{\Delta-\tfrac{d}2}I_{\Delta-\tfrac{d}2}\,,
\end{eqnarray}
as well as the $\Phi(z\to0)\sim z^{d-\Delta}$ modes
\begin{eqnarray}
z^{d-\Delta}\;_0F_1\!\left(;\tfrac{d}2-\Delta+1; \tfrac{z^2k^2}{4}\right)
=\Gamma\!\left(\tfrac{d}2-\Delta+1\right)z^{\tfrac{d}2}\left(\frac{k}2\right)^{\Delta-\tfrac{d}2}I_{\tfrac{d}2-\Delta}\,.
\end{eqnarray}
In the following, let us derive the Fourier transformation of them respectively.
\subsection{$\Phi(z\to0)\sim z^{\Delta}$ modes}
The Fourier transformation of the $\Phi(z\to0)\sim z^{\Delta}$ mode is
\begin{eqnarray}
\!\!\!\!&&\!\!\!\!\lim_{\Lambda\rightarrow\infty}\frac{z^{\frac{d}2}}{(2\pi)^{d}}
\int\rd^dk\,\ep^{-\frac{(zk)^2}{\Lambda^2}}I_{\Delta-\tfrac{d}2}\left(z k\right)
k^{{\Delta}-\tfrac{d}2}\ep^{\ii k(x-x')}
\cr\!\!\!\!&=&\!\!\!\!\lim_{\Lambda\rightarrow\infty}\frac{z^{\frac{d}2}}{(2\pi)^{d}}
\frac{2\pi^{\frac{d-1}{2}}}{\Gamma\!\left(\frac{d-1}{2}\right)}
\frac{\sqrt{\pi}\Gamma\!\left(\frac{d-1}{2}\right)}{\Gamma\!\left(\tfrac{d}2\right)}
\int_0^{\infty} \rd k\,\ep^{-\frac{(zk)^2}{\Lambda^2}}I_{\Delta-\tfrac{d}2}\left(z k\right)k^{\Delta+\tfrac{d}2-1}\,{}_0F_1\!\left(;\tfrac{d}2;-\tfrac{k^2r^2}{4}\right)
\cr\!\!\!\!&=&\!\!\!\!\lim_{\Lambda\rightarrow\infty}\frac{z^{\frac{d}2}}{(2\pi)^{d}}
\frac{2\pi^{\frac{d-1}{2}}}{\Gamma\!\left(\frac{d-1}{2}\right)}
\frac{\sqrt{\pi}\Gamma\!\left(\frac{d-1}{2}\right)}{\Gamma\!\left(\tfrac{d}2\right)}
\cr&&\sum_{m=0}^{\infty}\frac{1}{m!\,\Gamma\!\left(\Delta+m-\tfrac{d}2+1\right)}\int_0^{\infty} \rd k\,\ep^{-\frac{(zk)^2}{\Lambda^2}}\left(\frac{zk}{2}\right)^{2m+\Delta-\tfrac{d}2}k^{\Delta+\tfrac{d}2-1}
\,{}_0F_1\!\left(;\tfrac{d}2;-\tfrac{k^2r^2}{4}\right)
\cr\!\!\!\!&=&\!\!\!\!\lim_{\Lambda\rightarrow\infty}\frac{1}{(2\pi)^{d}}
\frac{2\pi^{\frac{d-1}{2}}}{\Gamma\!\left(\frac{d-1}{2}\right)}
\frac{\sqrt{\pi}\Gamma\!\left(\frac{d-1}{2}\right)}{2\,\Gamma\!\left(\tfrac{d}2\right)}
\cr&&\sum_{m=0}^{\infty}\frac{2^{-2 m-\Delta+\tfrac{d}2}z^{-\Delta}}{m!\,\Gamma\!\left(\Delta+m-\tfrac{d}2+1\right)}
\Gamma(\Delta+m)\Lambda^{2(\Delta+m)}{}_1F_1\!\left(\Delta+m;\tfrac{d}2;-\tfrac{\Lambda^2r^2}{4z^2}\right)
\end{eqnarray}
where we have used the fact that the convergence radius of ${}_0F_1$ and ${}_1F_1$ are $\infty$ and
\begin{eqnarray}
\int_0^{\pi}\rd\theta\,(\sin{\theta})^{d-2}\,\ep^{\ii kr\cos\theta}
\!\!\!\!&=&\!\!\!\!\frac{\sqrt{\pi}\Gamma\!\left(\tfrac{d-1}{2}\right)}{\Gamma\!\left(\tfrac{d}2\right)}{}_0F_1\!\left(;\tfrac{d}2;-\tfrac{k^2r^2}{4}\right)\,,
\cr
\int_0^{\infty} \rd k\,\ep^{-\frac{k^2}{\Lambda^2}}k^{2\Delta-1+2n}\!\!\!\!&=&\!\!\!\!\frac{1}{2}\Lambda^{2(\Delta+n)}\,\Gamma(\Delta+{n})~~~~~(\Delta+n>0)\,.\nn
\end{eqnarray}
By using the asymptotic expansion of confluent hypergeometric function
\begin{eqnarray}\label{asymp}
{}_1F_1(\alpha;\gamma;z)
\!\!\!\!&=&\!\!\!\!\frac{\Gamma(\gamma)}{\Gamma(\gamma-\alpha)}\ep^{\ii\pi \alpha}z^{-\alpha}{}_2F_0\!\left(\alpha,\alpha-\gamma+1;;-z^{-1}\right)
\cr&&+\frac{\Gamma(\gamma)}{\Gamma(\alpha)}\ep^{z}z^{\alpha-\gamma}{}_2F_0\!\left(\gamma-\alpha,1-\alpha;;z^{-1}\right)
~~~~{\rm for}~~~-\frac{\pi}{2}<\arg z<\frac{3\pi}{2}\nn
\end{eqnarray}
we find in the regime $(x-x')^2>z^2$
\begin{eqnarray}
\!\!\!\!&&\!\!\!\!\lim_{\Lambda\rightarrow\infty}\frac{1}{(2\pi)^{d}}
\frac{2\pi^{\frac{d-1}{2}}}{\Gamma\!\left(\frac{d-1}{2}\right)}
\frac{\sqrt{\pi}\Gamma\!\left(\frac{d-1}{2}\right)}{2\,\Gamma\!\left(\tfrac{d}2\right)}
\sum_{m=0}^{\infty}\frac{2^{-2 m-\Delta+\tfrac{d}2}z^{-\Delta}}{m!\,\Gamma\!\left(\Delta+m-\tfrac{d}2+1\right)}
\Gamma(\Delta+m)\Lambda^{2(\Delta+m)}{}_1F_1\!\left(\Delta+m;\tfrac{d}2;-\frac{\Lambda^2r^2}{4z^2}\right)
\cr\!\!\!\!&=&\!\!\!\!
\frac{2^{\Delta-\frac{d}2}z^{-\Delta}\sin\left(\frac{(d-2\Delta)\pi}{2}\right)\,\Gamma(\Delta)}
{\pi^{\frac{d}2+1}}
\sum_{m=0}^{\infty}
\frac{(-1)^m\,\Gamma(\Delta+m)}{m!\,\Gamma(\Delta)}
\frac{z^{2(\Delta+m)}}{(x-x')^{2(\Delta+m)}}
\cr\!\!\!\!&=&\!\!\!\!
\frac{2^{\Delta-\frac{d}2}\sin\left(\frac{(d-2\Delta)\pi}{2}\right)\,\Gamma(\Delta)}
{\pi^{\frac{d}2+1}}
\left(\frac{z}{z^2+(x-x')^2}\right)^{\Delta}\,.
\end{eqnarray}
Thus the $\Phi(z\to0)\sim z^{\Delta}$ mode reproduces $K(z,x;x')$ for the region $(x-x')^2>z^2$ as expected.
\\For $(x-x')^2<z^2$, we exchange the order of summations and find
\begin{eqnarray}\label{ndiv}
\!\!\!\!&&\!\!\!\!\lim_{\Lambda\rightarrow\infty}\frac{1}{(2\pi)^{d}}
\frac{2\pi^{\frac{d-1}{2}}}{\Gamma\!\left(\frac{d-1}{2}\right)}
\frac{\sqrt{\pi}\Gamma\!\left(\frac{d-1}{2}\right)}{2\,\Gamma\!\left(\tfrac{d}2\right)}
\sum_{m=0}^{\infty}\frac{2^{-2 m-\Delta+\tfrac{d}2}z^{-\Delta}}{m!\,\Gamma\!\left(\Delta+m-\tfrac{d}2+1\right)}
\Gamma(\Delta+m)\Lambda^{2(\Delta+m)}{}_1F_1\!\left(\Delta+m;\tfrac{d}2;-\frac{\Lambda^2r^2}{4z^2}\right)
\cr\!\!\!\!&=&\!\!\!\!\lim_{\Lambda\rightarrow\infty}\frac{1}{(2\pi)^{d}}
\frac{2\pi^{\frac{d-1}{2}}}{\Gamma\!\left(\frac{d-1}{2}\right)}
\frac{\sqrt{\pi}\Gamma\!\left(\frac{d-1}{2}\right)}{2\,\Gamma\!\left(\tfrac{d}2\right)}
\sum_{m=0}^{\infty}\sum_{n=0}^{\infty}\frac{2^{-2 m-\Delta+\tfrac{d}2}z^{-\Delta}}{m!\,\Gamma\!\left(\Delta+m-\tfrac{d}2+1\right)}
\frac{\Gamma(\Delta+m+n)\Gamma\!\left(\tfrac{d}2\right)}{n!\,\Gamma\!\left(\tfrac{d}2+n\right)}
\Lambda^{2(\Delta+m)}\left(-\frac{\Lambda^2r^2}{4z^2}\right)^n
\cr\!\!\!\!&=&\!\!\!\!\lim_{\Lambda\rightarrow\infty}\frac{1}{(2\pi)^{d}}
\frac{2\pi^{\frac{d-1}{2}}}{\Gamma\!\left(\frac{d-1}{2}\right)}
\frac{\sqrt{\pi}\Gamma\!\left(\frac{d-1}{2}\right)}{2\,\Gamma\!\left(\tfrac{d}2\right)}
\cr&&\sum_{n=0}^{\infty}\frac{2^{-\Delta+\tfrac{d}2}\Lambda^{2\Delta}z^{-\Delta}}{\Gamma\!\left(\Delta-\tfrac{d}2+1\right)}
\frac{\Gamma(\Delta+n)\Gamma\!\left(\tfrac{d}2\right)}{n!\,\Gamma\!\left(\tfrac{d}2+n\right)}
{}_1F_1\!\left(\Delta+n;\Delta-\tfrac{d}2+1;\frac{\Lambda^{2}}{4}\right)\left(-\frac{\Lambda^2r^2}{4z^2}\right)^n
\cr\!\!\!\!&=&\!\!\!\!\frac{1}{(2\pi)^{d}}
\frac{2\pi^{\frac{d-1}{2}}}{\Gamma\!\left(\frac{d-1}{2}\right)}
\frac{\sqrt{\pi}\Gamma\!\left(\frac{d-1}{2}\right)}{2\,\Gamma\!\left(\tfrac{d}2\right)}
\frac{2^{\Delta+\tfrac{d}2}z^{-\Delta}\Gamma(\Delta)
\sin\left(\frac{d\pi}{2}\right)\Gamma\!\left(\tfrac{d}2\right)}{\pi}
\sum_{n=0}^{\infty}\frac{\Gamma(\Delta+n)}{n!\,\Gamma(\Delta)}
\left(-\frac{r^2}{z^2}\right)^n
\cr&&
\left[\ep^{\ii\pi\Delta}
+\lim_{\Lambda\rightarrow\infty}(-1)^n\frac{\Gamma\!\left(1-\tfrac{d}2-n\right)}{\Gamma\!\left(\Delta+n\right)}\ep^{\frac{\Lambda^{2}}{4}}\left(\frac{\Lambda^{2}}{4}\right)^{\Delta+\tfrac{d}2+2n-1}
{}_2F_0\!\left(1-\tfrac{d}2-n;1-\Delta-n;\frac{4}{\Lambda^{2}}\right)\right]
\cr\!\!\!\!&=&\!\!\!\!
\frac{2^{\Delta-\frac{d}2}\Gamma(\Delta)}{\pi^{\frac{d}2+1}}
\ep^{\ii\pi\Delta}\sin\left(\tfrac{d\pi}{2}\right)
\left(\frac{z}{z^2+(x-x')^2}\right)^{\Delta}
\cr\!\!\!\!&&\!\!\!\!
+\frac{2^{\Delta-\frac{d}2}}
{\pi^{\frac{d}2}}\lim_{\Lambda\rightarrow\infty}\sum_{n=0}^{\infty}
\frac{\left(-\frac{r^2}{z^2}\right)^n}{n!\,\Gamma\!\left(\tfrac{d}2+n\right)}
 \ep^{\frac{\Lambda^{2}}{4}}\left(\frac{\Lambda^{2}}{4}\right)^{\Delta+\tfrac{d}2+2n-1}
{}_2F_0\!\left(1-\tfrac{d}2-n;1-\Delta-n;\frac{4}{\Lambda^{2}}\right)
\end{eqnarray}
The result is divergent and the divergent parts are expressed as terms with positive powers of $\Lambda$.
\subsection{$\Phi(z\to0)\sim z^{d-\Delta}$ modes}
For the $\Phi(z\to0)\sim z^{\Delta}$ modes, in the regime $(x-x')^2>z^2$, we have
\begin{eqnarray}
\!\!\!\!&&\!\!\!\!\lim_{\Lambda\rightarrow\infty}\frac{z^{\frac{d}2}}{(2\pi)^{d}}
\int\rd^dk\,\ep^{-\frac{(zk)^2}{\Lambda^2}}I_{\frac{d}2-\Delta}\left(z k\right)
k^{{\Delta}-\tfrac{d}2}\ep^{\ii k(x-x')}
\cr\!\!\!\!&=&\!\!\!\!\lim_{\Lambda\rightarrow\infty}\frac{z^{\frac{d}2}}{(2\pi)^{d}}
\frac{2\pi^{\frac{d-1}{2}}}{\Gamma\!\left(\frac{d-1}{2}\right)}
\frac{\sqrt{\pi}\Gamma\!\left(\frac{d-1}{2}\right)}{\Gamma\!\left(\tfrac{d}2\right)}
\int_0^{\infty} \rd k\,\ep^{-\frac{(zk)^2}{\Lambda^2}}I_{\frac{d}2-\Delta}\left(z k\right)k^{\Delta+\tfrac{d}2-1}\,{}_0F_1\!\left(;\tfrac{d}2;-\frac{k^2r^2}{4}\right)
\cr\!\!\!\!&=&\!\!\!\!\lim_{\Lambda\rightarrow\infty}\frac{z^{\frac{d}2}}{(2\pi)^{d}}
\frac{2\pi^{\frac{d-1}{2}}}{\Gamma\!\left(\frac{d-1}{2}\right)}
\frac{\sqrt{\pi}\Gamma\!\left(\frac{d-1}{2}\right)}{\Gamma\!\left(\tfrac{d}2\right)}
\cr&&\sum_{m=0}^{\infty}\frac{1}{m!\,\Gamma\!\left(-\Delta+m+\tfrac{d}2+1\right)}\int_0^{\infty} \rd k\,\ep^{-\frac{(zk)^2}{\Lambda^2}}\left(\frac{zk}{2}\right)^{2m-\Delta+\tfrac{d}2}k^{\Delta+\tfrac{d}2-1}\,{}_0F_1\!\left(;\tfrac{d}2;-\frac{k^2r^2}{4}\right)
\cr\!\!\!\!&=&\!\!\!\!\lim_{\Lambda\rightarrow\infty}\frac{1}{(2\pi)^{d}}
\frac{2\pi^{\frac{d-1}{2}}}{\Gamma\!\left(\frac{d-1}{2}\right)}
\frac{\sqrt{\pi}\Gamma\!\left(\frac{d-1}{2}\right)}{2\,\Gamma\!\left(\tfrac{d}2\right)}
\sum_{m=0}^{\infty}\frac{2^{-2 m+\Delta-\tfrac{d}2}z^{-\Delta}}{m!\,\Gamma\!\left(-\Delta+m+\tfrac{d}2+1\right)}
\Gamma\!\left(\tfrac{d}2+m\right)\Lambda^{d+2m}{}_1F_1\!\left(\tfrac{d}2+m;\tfrac{d}2;-\frac{\Lambda^2r^2}{4z^2}\right)
\cr\!\!\!\!&=&\!\!\!\!
\sum_{m=0}^{\infty}\frac{2^{-2 m+\Delta-\tfrac{d}2}z^{-\Delta}}{m!\,\Gamma\!\left(-\Delta+m+\tfrac{d}2+1\right)}
\frac{2^{m}\,\Gamma\!\left(\tfrac{d}2+m\right)}
{\pi^{\frac{d}2}\,\Gamma\!\left(-m\right)}
\frac{z^{d+2m}}{(x-x')^{d+2m}}
=0\,.
\end{eqnarray}
where we have also used the asymptotic expansion (\ref{asymp}) of confluent hypergeometric function ${}_1F_1$.
The above result shows that the $\Phi(z\to0)\sim z^{d-\Delta}$ mode is a quasi-local function which is identically vanishing when $(x-x')^2>z^2$.
\\In the $(x-x')^2<z^2$ regime, we have
\begin{eqnarray}\label{nndiv}
\!\!\!\!&&\!\!\!\!\frac{z^{\frac{d}2}}{(2\pi)^{d}}
\int\rd^dk\,\ep^{-\frac{(zk)^2}{\Lambda^2}}I_{\frac{d}2-\Delta}\left(z k\right)
k^{{\Delta}-\tfrac{d}2}\ep^{\ii k(x-x')}
\cr\!\!\!\!&=&\!\!\!\!
\frac{2^{\Delta-\frac{d}2}z^{d-\Delta}}
{\,\Gamma\!\left(\tfrac{d}2-\Delta+1\right)}
{}_0F_1\!\left(;\tfrac{d}2-\Delta+1;-\tfrac{z^2}{4}\Box\right)
\delta^d(x-x')
\cr\!\!\!\!&=&\!\!\!\!\lim_{\Lambda\rightarrow\infty}\frac{2^{\Delta-\frac{d}2}z^{d-\Delta}}
{\Gamma\!\left(\tfrac{d}2-\Delta+1\right)}
{}_0F_1\!\left(;\tfrac{d}2-\Delta+1;-\tfrac{z^2}{4}\Box\right)
\frac{\Lambda^d}{2^d\pi^{\frac{d}2}z^d}
\ep^{-\frac{(x-x')^2\Lambda^2}{4z^2}}
\cr\!\!\!\!&=&\!\!\!\!\lim_{\Lambda\rightarrow\infty}\frac{2^{\Delta-\frac{d}2}z^{d-\Delta}}
{\,\Gamma\!\left(\tfrac{d}2-\Delta+1\right)}\frac{\Lambda^d}{2^d\pi^{\frac{d}2}z^d}
{}_0F_1\!\left(;\tfrac{d}2-\Delta+1;-\tfrac{z^2}{4}\Box\right)
\sum_{m=0}^{\infty}\frac{1}{m!}\left(\frac{-(x-x')^2\Lambda^2}{4z^2}\right)^m
\cr\!\!\!\!&=&\!\!\!\!\lim_{\Lambda\rightarrow\infty}\frac{2^{\Delta-\frac{d}2}z^{d-\Delta}}
{\,\Gamma\!\left(\tfrac{d}2-\Delta+1\right)}\frac{\Lambda^d}{2^d\pi^{\frac{d}2}z^d}
\cr&&\sum_{m=0}^{\infty}\sum_{n=0}^{m}
\frac{\Gamma\!\left(\tfrac{d}2-\Delta+1\right)}{n!\,\Gamma\!\left(\tfrac{d}2-\Delta+1+n\right)}
\frac{\Gamma\!\left(\tfrac{d}2+m\right)}{\Gamma(m-n+1)\,\Gamma\!\left(\tfrac{d}2+m-n\right)}
\left(\frac{\Lambda^2}{4}\right)^m
\left(-\frac{(x-x')^2}{z^2}\right)^{(m-n)}
\cr\!\!\!\!&=&\!\!\!\!
\lim_{\Lambda\rightarrow\infty}\frac{2^{\Delta-\frac{d}2}z^{d-\Delta}}
{\Gamma\!\left(\tfrac{d}2-\Delta+1\right)}\frac{\Lambda^d}{2^d\pi^{\frac{d}2}z^d}
\sum_{k=0}^{\infty}
\frac{1}{k!}
{}_1F_1\!\left(\tfrac{d}2+k;\tfrac{d}2-\Delta+1;\frac{\Lambda^2}{4}\right)
\left(-\frac{\Lambda^2(x-x')^2}{4z^2}\right)^{k}
\cr\!\!\!\!&=&\!\!\!\!
\frac{2^{\Delta-\frac{d}2}z^{-\Delta}}
{\pi^{\frac{d}2}}
\sum_{k=0}^{\infty}
\left[\ep^{\ii\frac{d\pi}{2}}\frac{\sin(\pi\Delta)}{\pi}\Gamma(\Delta)\frac{\Gamma\!\left(\Delta+k\right)}{k!\,\Gamma(\Delta)}
\left(-\frac{(x-x')^2}{z^2}\right)^{k}
\right.\cr&&~\left.
+\lim_{\Lambda\rightarrow\infty}\frac{\left(-\frac{(x-x')^2}{z^2}\right)^{k}}{k!\,\Gamma\!\left(\frac{d}2+k\right)}
\ep^{\frac{\Lambda^{2}}{4}}\left(\frac{\Lambda^2}{4}\right)^{\Delta+\frac{d}2+2k-1}
{}_2F_0\!\left(1-\Delta-k,1-\tfrac{d}2-k;;\frac{4}{\Lambda^2}\right)\right]
\cr\!\!\!\!&=&\!\!\!\!
\frac{2^{\Delta-\frac{d}2}\,\Gamma(\Delta)}
{\pi^{\frac{d}2+1}}\ep^{\ii\frac{d\pi}{2}}\,\sin(\pi\Delta)
\left(\frac{z}{z^2+(x-x')^2}\right)^{\Delta}
\cr\!\!\!\!&&\!\!\!\!
+\lim_{\Lambda\rightarrow\infty}\frac{2^{\Delta-\frac{d}2}z^{-\Delta}}
{\pi^{\frac{d}2}}
\sum_{k=0}^{\infty}\frac{\left(-\frac{(x-x')^2}{z^2}\right)^{k}}{k!\,\Gamma\!\left(\frac{d}2+k\right)}
\ep^{\frac{\Lambda^{2}}{4}}\left(\frac{\Lambda^2}{4}\right)^{\Delta+\frac{d}2+2k-1}
{}_2F_0\!\left(1-\Delta-k,1-\tfrac{d}2-k;;\frac{4}{\Lambda^2}\right)\,.~~~~~~~~~~~~~~
\end{eqnarray}
As in the $\Phi(z\to0)\sim z^{\Delta}$ case, the result is divergent and the divergent parts are expressed as terms with positive powers of $\Lambda$.

Comparing (\ref{ndiv}) and (\ref{nndiv}), we notice that the divergence in the region $(x-x')^2<z^2$ can only  be canceled in the combination $$I_{\Delta-\frac{d}2}-I_{\frac{d}2-\Delta}\sim K_{\Delta-\frac{d}2},$$
which is coincide with the answer obtained in \cite{Balasubramanian:1998sn} by requiring the $z\to\infty$ regularity of the momentum space solution.
We also notice that the remnant finite term gives rise to exactly the analytical continuation of the result in $(x-x')^2>z^2$ region. Thus one can take the continuity at $|x-x'|^2=z^2$ as the prescription of the regularization in the coordinate space.

The above derivations are performed under the Euclidean signature. For the Minkowskian signature, the 2-point propagator is not unique due to the existence of the lightcone singularity. Depending on which kind of 2-point propagator was considered, the corresponding bulk momentum space formulae are different.  These different formulae are related to the different choices of quantum states of the boundary QFT \cite{Balasubramanian:1998de,Skenderis:2008dg}.
For example, as pointed in \cite{Son:2002sd,Marolf:2004fy,Iqbal:2009fd}, the retarded propagator is related to take the ingoing boundary condition at the horizon $z\to\infty$; while the advanced propagator is related to take the out-coming boundary condition at the horizon $z\to\infty$. In these two cases, the relevant bulk  momentum space formulae are still linear combinations of the $\Phi(z\to0)\sim z^{\Delta}$ mode and the $\Phi(z\to0)\sim z^{d-\Delta}$ mode. It is also possible to have the  bulk  momentum space formulae with purely normalizable modes where $\Phi(z\to0)\sim z^{\Delta}$ and $k^2<0$. This case is in fact the one been discussed in \cite{Banks:1998dd,Balasubramanian:1999ri,Bena:1999jv}.
In our present consideration, the formula is related to the radial quantization which is initially well established under the Euclidean signature. Therefore, the relevant bulk momentum space formula is taken to be the simple analytical continuation of the Euclidean one \cite{Balasubramanian:1998sn} which is uniquely fixed by requiring the regularity at $z\to\infty$ since $k^2>0$ for the Euclidean signature.


\begin{thebibliography}{99}

\bibitem{Maldacena:1997re}
  J.~M.~Maldacena,
  ``The Large N limit of superconformal field theories and supergravity'',
  Int.\ J.\ Theor.\ Phys.\  {\bf 38}, 1113 (1999)
  [Adv.\ Theor.\ Math.\ Phys.\  {\bf 2}, 231 (1998)]
  [hep-th/9711200].

\bibitem{Gubser:1998bc}
  S.~S.~Gubser, I.~R.~Klebanov and A.~M.~Polyakov,
  ``Gauge theory correlators from noncritical string theory'',
  Phys.\ Lett.\ B {\bf 428}, 105 (1998)
  [hep-th/9802109].

\bibitem{Witten:1998qj}
  E.~Witten,
  ``Anti-de Sitter space and holography'',
  Adv.\ Theor.\ Math.\ Phys.\  {\bf 2}, 253 (1998)
  [hep-th/9802150].

\bibitem{Banks:1998dd}
  T.~Banks, M.~R.~Douglas, G.~T.~Horowitz and E.~J.~Martinec,
  ``AdS dynamics from conformal field theory'',
  hep-th/9808016.

\bibitem{Balasubramanian:1999ri}
  V.~Balasubramanian, S.~B.~Giddings and A.~E.~Lawrence,
  ``What do CFTs tell us about Anti-de Sitter space-times?'',
  JHEP {\bf 9903}, 001 (1999)
  [hep-th/9902052].


\bibitem{Bena:1999jv}
  I.~Bena,
  ``On the construction of local fields in the bulk of AdS(5) and other spaces'',
  Phys.\ Rev.\ D {\bf 62}, 066007 (2000)
  [hep-th/9905186].



\bibitem{Hamilton:2006az}
  A.~Hamilton, D.~N.~Kabat, G.~Lifschytz and D.~A.~Lowe,
  ``Holographic representation of local bulk operators'',
  Phys.\ Rev.\ D {\bf 74}, 066009 (2006)
  [hep-th/0606141];
  A.~Hamilton, D.~N.~Kabat, G.~Lifschytz and D.~A.~Lowe,
  ``Local bulk operators in AdS/CFT: A Holographic description of the black hole interior'',
  Phys.\ Rev.\ D {\bf 75}, 106001 (2007)
  [Phys.\ Rev.\ D {\bf 75}, 129902 (2007)]
  [hep-th/0612053].

\bibitem{Susskind:1998dq}
  L.~Susskind and E.~Witten,
  ``The Holographic bound in anti-de Sitter space'',
  hep-th/9805114.

\bibitem{Balasubramanian:1998sn}
  V.~Balasubramanian, P.~Kraus and A.~E.~Lawrence,
  ``Bulk versus boundary dynamics in anti-de Sitter space-time'',
  Phys.\ Rev.\ D {\bf 59}, 046003 (1999)
  [hep-th/9805171].

\bibitem{Freedman:1998tz}
  D.~Z.~Freedman, S.~D.~Mathur, A.~Matusis and L.~Rastelli,
  ``Correlation functions in the CFT(d) / AdS(d+1) correspondence'',
  Nucl.\ Phys.\ B {\bf 546}, 96 (1999)
  [hep-th/9804058].

\bibitem{Pappadopulo:2012jk}
  D.~Pappadopulo, S.~Rychkov, J.~Espin and R.~Rattazzi,
  ``OPE Convergence in Conformal Field Theory'',
  Phys.\ Rev.\ D {\bf 86}, 105043 (2012)
  [arXiv:1208.6449 [hep-th]];
  ~S.~Rychkov, EPFL Lectures on Conformal Field Theory in $D\geq3$ Dimensions [arXiv:1601.05000 [hep-th]].


\bibitem{Kabat:2011rz}
  D.~Kabat, G.~Lifschytz and D.~A.~Lowe,
  ``Constructing local bulk observables in interacting AdS/CFT'',
  Phys.\ Rev.\ D {\bf 83}, 106009 (2011)
  [arXiv:1102.2910 [hep-th]].


\bibitem{zlwi013}
Z.-L. Wang, Talk preseneted on KIAS-YITP Joint Workshop 2015: Geometry in Gauge Theories and String Theory.

\bibitem{Belavin:1984vu}
  A.~A.~Belavin, A.~M.~Polyakov and A.~B.~Zamolodchikov,
  ``Infinite Conformal Symmetry in Two-Dimensional Quantum Field Theory'',
  Nucl.\ Phys.\ B {\bf 241}, 333 (1984).

\bibitem{Fradkin:1996is}
  E.~S.~Fradkin and M.~Y.~Palchik,
  ``Conformal Quantum Field Theory in $D$-dimensions'',
  Kluwer Academic Publishers, 1996.

\bibitem{Fradkin:1997df}
  E.~S.~Fradkin and M.~Y.~Palchik,
  ``New developments in D-dimensional conformal quantum field theory'',
  Phys.\ Rept.\  {\bf 300}, 1 (1998).

\bibitem{Almheiri:2012rt}
  A.~Almheiri, D.~Marolf, J.~Polchinski and J.~Sully,
  ``Black Holes: Complementarity or Firewalls?,''
  JHEP {\bf 1302}, 062 (2013)
  [arXiv:1207.3123 [hep-th]].

\bibitem{Braunstein:2009my}
  S.~L.~Braunstein, S.~Pirandola and K.~\.Zyczkowski,
  ``Better Late than Never: Information Retrieval from Black Holes'',
  Phys.\ Rev.\ Lett.\  {\bf 110}, no. 10, 101301 (2013)
  [arXiv:0907.1190 [quant-ph]].

\bibitem{Nakayama:2015mva}
  Y.~Nakayama and H.~Ooguri,
  ``Bulk Locality and Boundary Creating Operators'',
  arXiv:1507.04130 [hep-th].

\bibitem{Miyaji:2015fia}
  M.~Miyaji, T.~Numasawa, N.~Shiba, T.~Takayanagi and K.~Watanabe,
  ``cMERA as Surface/State Correspondence in AdS/CFT'',
  arXiv:1506.01353 [hep-th].

\bibitem{Heemskerk:2010hk}
  I.~Heemskerk and J.~Polchinski,
  ``Holographic and Wilsonian Renormalization Groups'',
  JHEP {\bf 1106}, 031 (2011)
  [arXiv:1010.1264 [hep-th]].

\bibitem{Sarkar:2014jia}
  D.~Sarkar,
  ``(A)dS holography with a cutoff'',
  Phys.\ Rev.\ D {\bf 90}, no. 8, 086005 (2014)
  [arXiv:1408.0415 [hep-th]].

\bibitem{Balasubramanian:1998de}
  V.~Balasubramanian, P.~Kraus, A.~E.~Lawrence and S.~P.~Trivedi,
  ``Holographic probes of anti-de Sitter space-times'',
  Phys.\ Rev.\ D {\bf 59} (1999) 104021
  [hep-th/9808017].

\bibitem{Skenderis:2008dg}
  K.~Skenderis and B.~C.~van Rees,
  ``Real-time gauge/gravity duality: Prescription, Renormalization and Examples,''
  JHEP {\bf 0905} (2009) 085
  [arXiv:0812.2909 [hep-th]].

\bibitem{Son:2002sd}
  D.~T.~Son and A.~O.~Starinets,
  ``Minkowski space correlators in AdS / CFT correspondence: Recipe and applications'',
  JHEP {\bf 0209} (2002) 042
  [hep-th/0205051].

\bibitem{Marolf:2004fy}
  D.~Marolf,
  ``States and boundary terms: Subtleties of Lorentzian AdS / CFT'',
  JHEP {\bf 0505} (2005) 042
  [hep-th/0412032].

\bibitem{Iqbal:2009fd}
  N.~Iqbal and H.~Liu,
  ``Real-time response in AdS/CFT with application to spinors'',
  Fortsch.\ Phys.\  {\bf 57} (2009) 367
  [arXiv:0903.2596 [hep-th]].
\end{thebibliography}
\end{document}